# Questioning the reliability of estimates of enzyme inhibitor constant: Case of competitive inhibition

Sharmistha Dhatt

Department of Chemistry, University of Calcutta, Rajabazar Science College Campus

92, A. P. C. Road, Kolkata 700009, India.

**Abstract**

Reliability of kinetic parameters are crucial in understanding enzyme kinetics within cellular system. The present study suggests a few cautions that need introspection for estimation of parameters like $K_M$, $V_{Max}$ and $K_I$ using Lineweaver-Burk plots. The quality of $IC_{50}$ too needs a thorough reinvestigation because of its direct link with $K_I$ and $K_M$ values. Inhibition kinetics under both steady-state and non-steady-state conditions are studied and errors in estimated parameters are compared against actual values to settle the question of their adequacy.

**List of abbreviations:**

MM: Michaelis and Menten
LB: Lineweaver-Burk
QSSA: Quasi-steady-state approximation
CI: Competitive inhibition

E-mail: pcsdhatt@gmail.com

## 1. Introduction

Since the publication of the seminal paper [1] of Michaelis and Menten (MM) on enzyme kinetics in 1913, molecular biology has evolved tremendously. Nevertheless, the importance of century-old MM kinetics has not faded away [2]. It is still used in several contexts. Inhibition kinetics of enzyme characterization in large biochemical network is one such area. Many drugs alter the activity of specific enzymes within our body. Therefore, enzyme inhibition studies are routinely conducted [3-4] to assess the presence and magnitude of drug-drug interaction. Estimating the inhibition constant ($K_I$) is particularly important in this regard. It offers a critical information for a specific drug. Another popular alternative measure is $IC_{50}$, the anagonistic drug potency, telling us the concentration of a drug that must be present in the body to achieve a desired degree of enzyme inhibition.

This article concentrates on competitive reversible inhibitors that interact with enzymes to alter either their substrate binding affinity, or catalytic activity, or both. Extracting values for MM parameters for a given enzyme so as to quantify the effect of inhibition is identified as a nonlinear optimization problem. Over the years, various methods [4-7] to determine the kinetic parameters like $K_M$, $V_{Max}$ and $K_I$ have been proposed in presence of inhibitors. The statistical limitation of such parametric estimations [8-11] is also known. Graphical estimates are, however, much more common. But, they do not provide any assessment of errors of calculated data. This include the most popular Lineweaver-Burk (LB) plot [2, 6], and even the advanced scheme of Eisenthan and Cornish-Bowden [5-7, 12-14]. This is where our article intends to strike a note. A further insight into the legitimacy of the quasi-steady state approximation (QSSA) [15-17] is checked in the context of errors in $K_I$ estimation through LB plots. Indeed, we notice minimum error in estimated values under QSSA conditions only, and that too under specific circumstances, to be pointed out in due course.

## 2. The scheme

In competitive inhibition (CI), substrate and inhibitor that bind to free enzyme are mutually exclusive. CI acts only to increase the MM constant for the substrate. The scheme is popularly designated as

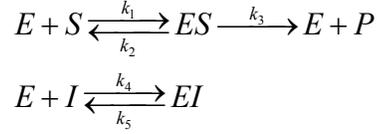

We employ the following dimensionless variables for study:

$$\alpha = \frac{E}{E_0}, \ \beta = \frac{S}{E_0}, \ \gamma = \frac{ES}{E_0}, \ \xi = \frac{I}{E_0}, \ \nu = \frac{EI}{E_0}, \ \delta = \frac{P}{E_0}, \ \tau = k_3 t \quad (1)$$

This scaling scheme has elsewhere [16] been found to be quite convenient, though other possibilities exist [12, 14]. The coupled kinetic equations that follow from the scheme then reduce to

$$\frac{d\beta}{d\tau} = -K_1 \alpha \beta + K_2 \gamma \quad (2)$$

$$\frac{d\alpha}{d\tau} = -K_1 \alpha \beta + (K_2 + 1)\gamma - K_3 \alpha \xi - K_4 \nu \quad (3)$$

$$\frac{d\gamma}{d\tau} = K_1 \alpha \beta - (K_2 + 1)\gamma \quad (4)$$

$$\frac{d\xi}{d\tau} = -K_3 \alpha \xi + K_4 \nu = -\frac{d\nu}{d\tau} \quad (5)$$

The constants $K_i$ (i = 1, ..., 4) are given by

$$K_1 = k_1 E_0 / k_3, \ K_2 = k_2 / k_3, \ K_3 = k_4 E_0 / k_3, \ K_4 = k_5 / k_3 \quad (6)$$

In addition, we have three mass-conservation equations

$$\alpha_0 = \alpha + \gamma + \nu; \ \xi_0 = \xi + \nu; \ \beta_0 = \beta + \gamma + \delta \quad (7)$$

The kinetic pathway is simulated numerically over time. Simulation does not presuppose QSSA for the catalytic path or rapid-equilibrium for the inhibitory route, as often assumed while elucidating competitive inhibition kinetics. Numerical simulation shows the temporal profile of the complex γ passing through a maximum, and the corresponding time is referred to as $\tau_C$. This time $\tau_C$ is a measure of the 'transient phase'. For convenience, the scaled concentrations at $\tau_C$ are designated by $\beta_C$, $\gamma_C$ and $\xi_C$. The appropriate Lineweaver-Burk equation [16] for the kinetics under such condition is

$$\frac{1}{\gamma_c} = \frac{K_M}{\beta_C [1 - (\xi_0 - \xi_C)]} + \frac{1}{[1 - (\xi_0 - \xi_C)]} \quad (8)$$

Equation (8), however, assumes only that $d\gamma/d\tau = 0$ at $\tau = \tau_C$. If we additionally impose the rapid equilibrium condition, i.e., $dv/d\tau = 0$, and also insist that $\xi_C = \xi_0$, then (8) reduces to the well-accepted form

$$\frac{1}{\gamma_C} = 1 + \frac{K_M}{\beta_C}\left(1 + \frac{\xi_0}{K_I}\right) \tag{9}$$

Here, $K_M$ (in scaled form) in Equations. (8) and (9) is given by $(K_2 + 1)/K_1$. For brevity, we mention here that, in our scaled formulation, rate of reaction is proportional to the complex concentration. Hence, our LB plot takes $1/\gamma_C$ as the ordinate and shows the variation against $1/\beta_C$ [16].

Estimation of $K_I$ is concluded from the equation

$$\frac{K'_M}{K_M} = \left(1 + \frac{\xi_0}{K_I}\right) \tag{10}$$

where $K_M$ and $K'_M$ are MM constants in absence and presence of inhibitors, respectively. The potential enzyme inhibition of a drug is quantified [18-23] in terms of $IC_{50}$ value where reversible inhibition is a desired mode of action. Popular Cheng-Prusoff equation [19] is used for such estimation via the relation

$$IC_{50} = K_I\left(1 + \frac{S}{K_M}\right) \tag{11}$$

In equation (11), S denotes the particular substrate concentration at which $IC_{50}$ is calculated for an enzyme assay in a particular laboratory concentration.

**3. Results and discussion**

In our exploration, we choose two reported literature data of vital physiological relevance. System 1 [24] corresponds to an *in-vitro* kinetic inhibition study of acetylcholine esterase with galanthame as inhibitor, declared as a case of CI in Alzeheimer disease. Inhibitors of acetylcholine breakdown by acetylcholine esterase constitute the main therapeutic modality of Alzheimer's disease. System 2 data [9] refer to an in vitro interaction study of coumarin and pilocarpine with cytochrome $P_{450}$ inhibitor. The interest here lies in analyzing metabolic reactions catalyzed by human liver microsomes. This is also a CI case.

Table 1 summarizes the findings of our study for two different sets of reported predictor variables classified as System 1 and System 2. All concentrations are expressed in $\mu M$ unit. Our systems under study are characterized by the following values of kinetic parameters:

$$\text{System 1: } (K_M = 26.3,\ V_{Max} = 65.0,\ K_I = 3.4)$$

$$\text{System 2: } (K_M = 10.0,\ V_{Max} = 0.1,\ K_I = 5.0)$$

We further distinguish the systems as follows:

System 1A: k1 = 5.0E(+02), k2 = 11.80E(+03), k3 = 1.30E(+03), e0 = 5.0E(-02)

System 1B: k1 = 5.0E(-01), k2 = 6.65E(+0), k3 = 6.5E(+0), e0 = 1.0E(+01)

System 2A: k1 = 1.5E(+02), k2 = 1.49E(+02), k3 = 1.0E(+01), e0 = 1.0E(-02)

System 2B: k1 = 5.0E(-02), k2 = 4.0E(-01), k3 = 1.0E(-01), e0 = 1.0E(+0)

Table 1: Literature values ($K_{I,Lit}$) and calculated estimates ($K_{I,Cal}$) of $K_I$ for chosen rate constants at QSSA (1A, 2A) and non-QSSA (1B, 2B) conditions are shown, along with errors incurred.

| Systems | $E_0$ | Run | $K_4$ | $K_5$ | $K_{I,\ Lit}$ | $K_{I,\ Cal}$ | Error% |
|---|---|---|---|---|---|---|---|
| 1A | 0.05 | 1 | 10.0 | 34.0 | | 18.264 | 437.17 |
| | | 2 | 100.0 | 340.0 | 3.4 | 3.997 | 17.56 |
| | | 3 | 1000.0 | 3400.0 | | 3.451 | 1.2 |
| 1B | 10.0 | 1 | 0.1 | 0.34 | | 67.958 | 1898.76 |
| | | 2 | 1.0 | 3.4 | 3.4 | 10.043 | 95.38 |
| | | 3 | 10.0 | 34.0 | | 5.645 | 66.03 |
| 2A | 0.01 | 1 | 1.0 | 5.0 | | 12.567 | 151.34 |
| | | 2 | 10.0 | 50.0 | 5.0 | 6.97 | 39.4 |
| | | 3 | 100.0 | 500.0 | | 5.027 | 0.54 |
| 2B | 1.0 | 1 | 0.1 | 0.5 | | 9.64 | 92.8 |
| | | 2 | 1.0 | 5.0 | 5.0 | 7.69 | 53.8 |
| | | 3 | 10.0 | 50.0 | | 5.82 | 16.4 |

For a given set of reported $K_M$, $V_{Max}$ and $K_I$, the rate constants for the catalytic route and inhibitory path are constructed such that the scheme can be followed both under QSSA and non-QSSA regime. In the present study, A refers to the cases that obey the QSSA, while B represents situations under non-QSSA category.

With each of the above sets of parameters, simulation has been carried out with seven different scaled substrate concentrations (10-70) for a definite strength of inhibitor ($\xi 0 = 0.5$) and one control (i.e., no inhibitor). The specific rate constants ($k_4, k_5$) for the binding and dissociation of the inhibitor-enzyme complex are altered keeping the equilibrium ratio intact to ensure a gradual transition from slow to rapid equilibrium for the path. These correspond to different runs.

Another point of interest lies in the estimation of $IC_{50}$. Both $IC_{50}$ and $K_I$ are measures of an inhibitor's ability to block enzyme action. But, they are not equivalent. $K_I$ values are the intrinsic properties of particular enzyme-inhibitor pairs and provide more reliable estimates of inhibitor potency, in contrast to $IC_{50}$ data that vary with substrate concentration and laboratory enzyme assay conditions. However, in view of equation (11), errors in $IC_{50}$ would be similar to those in $K_I$.

Important observations from the above study are now listed below:

(i) Estimated $K_I$ is in good agreement with literature data for cases assessed under QSSA condition for the catalytic route and rapid equilibrium for the inhibitory path. System 1A (3) and 2A (3) reveal this feature transparently.

(ii) An appreciable rate of the inhibitory path is sufficient for a smaller error quote for QSSA, whereas, even rapid equilibrium is inadequate for error drop in the non-QSSA regime. A comparison between systems 1A (3) and 1B (3) or systems 2A (3) and 2B (3) would make it clear.

(iii) For non-QSSA cases, e.g., systems 1B and 2B (see Table 1), an initial reduction in error quote finally saturates to a value even if the order of rate constants for the inhibitory path are higher than the catalytic path. This implies, rapid equilibrium cannot ensure a worthy estimate of $K_I$ via linear regression.

(iv) LB plots for the test cases, are, however, all linear without raising any doubt on the erroneous predictions of data. For example, both the Figure 1 and Figure 2 show that the LB plot is linear irrespective of whether QSSA is valid or not, or error estimates in $K_I$ is large or small.

(v) QSSA and rapid equilibrium provide a conducive environment for precise data estimation, whereas non-QSSA, even with a rapid equilibrium condition, might not be a suitable criterion for accurate prediction of $K_I$. Another complimentary observation is the low variation in $\tau_C$ values for cases satisfying QSSA compared to those in the non-QSSA regime, as evident from Figure 3 and Figure 4. Approximate constancy in $\tau_C$ may be an indication of QSSA.

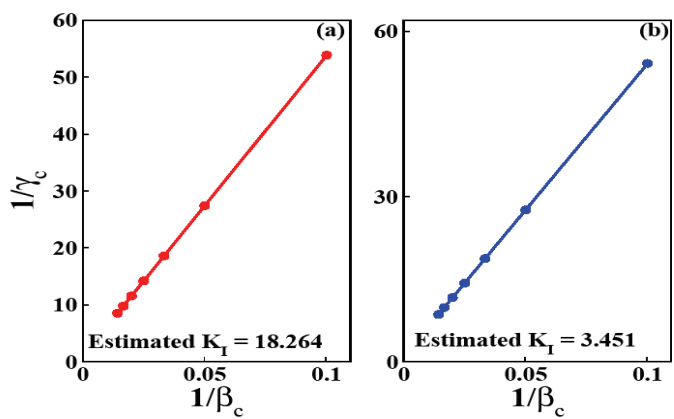

**Figure 1: Linear LB plot for System 1A (a) worst $K_I$ for 1A (1) (b) best $K_I$ for 1A (3).**

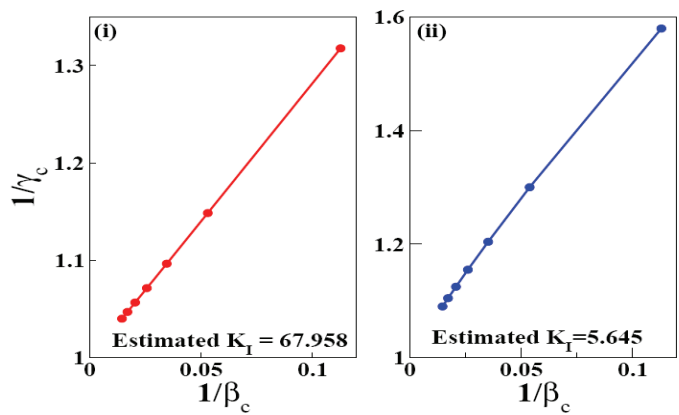

**Figure 2: Linear LB plot for System 1B (i) worst $K_I$ for 1B (1) (ii) best $K_I$ for 1B (3).**

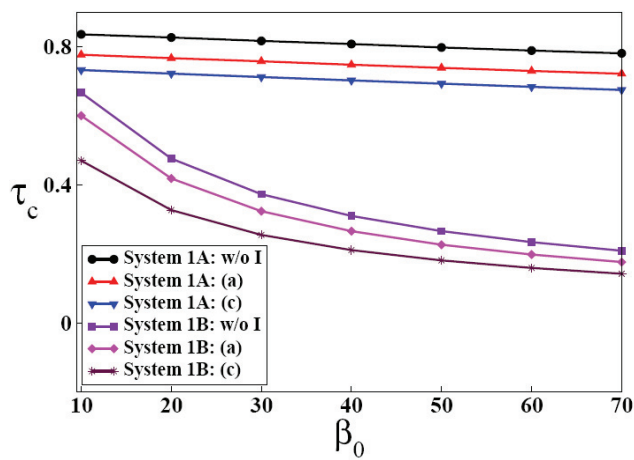

**Figure 3: Plot of $\tau_C$ vs substrate concentration ($\beta_0$) System 1A and 1B.**

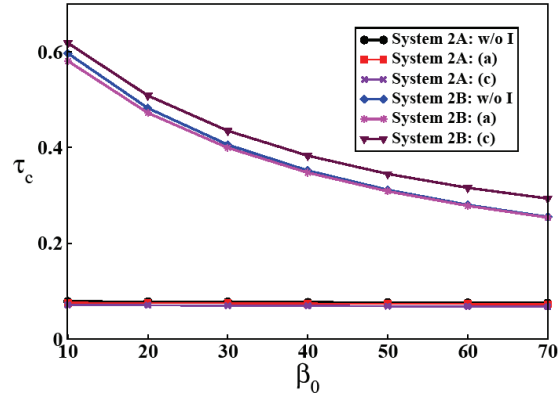

**Figure 4: Plot of $\tau_C$ vs substrate concentration ($\beta_0$) for System 2A and 2B.**

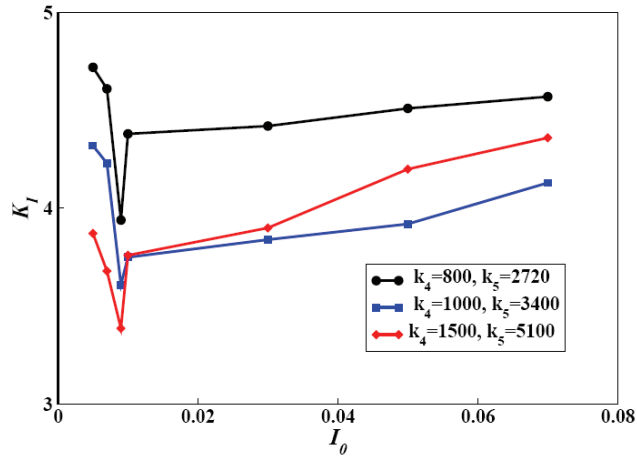

**Figure 5: Plot of $K_I$ vs $I_0$ for System 1A.**

(vi) The question of ensuring a rapid equilibrium between enzyme and inhibitor may be settled by varying the inhibitor concentration. Figure 5 sums up our observation for system 1A. The estimated $K_I$ passes through a minimum for a definite $I_0$ value, and also for a definite set of binding and dissociation constants, for the chosen inhibition path. Steep rise in the $K_I$ value with changing inhibitor concentration is an indication that one is on a wrong direction in data estimation. When the rate constants are altered, the minima in $K_I$ gradually saturate to the actual value, as shown in Figure 5. For any specific physiological process, however, the constraint lies in altering the biological rate constants. A bypass is to follow the variation with $I_0$ and report data at the minimum point.

**4. Conclusion**

The purpose of this study has been to put a caution on the rampant use of LB plot [2, 25-28] in analyzing in vitro enzyme inhibition data, subsequently obtaining $K_I$. The primary questions posed are (a) whether linearity of LB plot is enough to rely on estimated $K_I$ [29-30] without scrutinizing QSSA and non-QSSA regime, (b) whether reliability of such estimates depends on the adequacy of the QSSA, (c) whether an inhibitor ensuring rapid equilibrium condition provides better estimates of $K_I$ and (d) how to improve the error estimates and implement the method routinely for a robust estimate of $K_I$ minimizing the uncertainties involved therein.

Going through our analysis and computations for reversible CI, the following conclusions can be drawn:

(i) Situations where QSSA hold, and where inhibitors ensure rapid equilibrium condition, $K_I$ estimates are almost error free (see Table 1).

(ii) An observed linearity in LB plot may be quite misleading (see, e.g., Figures 1 and 2) in respect of the validity of QSSA, and hence the sanctity of $K_I$ values observed subsequently.

(iii) A virtual consistency of $\tau_C$ over large region of starting substrate concentration (see, e.g., Figures 3 and 4) is suggestive of whether QSSA holds for a particular case of enzyme-substrate-inhibitor combination under investigation.

(iv) The establishment of rapid equilibrium in the QSSA regime can be ensured by altering the inhibitor concentration to encounter a minimum in $K_I$. Lack of such a minimum (cf. Figure 5) will definitely warrant an attention for a new choice of enzyme-inhibitor pair.

In fine, we have illustrated a potential pitfall and remedial strategies for competitive inhibitors. The case of CI has been specifically chosen because it provides the most common pathway for drug-drug interactions. However, the same route can be routinely adopted for other inhibitor types as well. We hope, our effort will increase the awareness about errors incurred in $K_I$ (or $IC_{50}$) calculations, limiting their input in drug-dosage and drug-drug interactions in pharmaceutical industry. Some of our more interesting theoretical observations, displayed in Figures 3-5, may turn out to be useful guides to experimenters.

**Acknowledgement**

The author acknowledges CSIR, India for a research associateship, and thanks Dr. K. Banerjee and Prof. K. Bhattacharyya for valuable discussions and constructive comments.